\documentclass[twocolumn,aps,superscriptaddress]{revtex4}
\usepackage{amsmath, amssymb, verbatim, graphicx, color}

\def\e{{\rm e}}
\def\I{{\rm i}}
\newcommand{\eq}[1]{(\ref{#1})}
\newcommand{\ket}[1]{|#1\rangle}
\newcommand{\bra}[1]{\langle #1|}
\newcommand{\proj}[1]{|#1\rangle\!\langle #1|}

\newcommand{\beqa}{\begin{eqnarray}}
\newcommand{\eeqa}{\end{eqnarray}}
\newcommand{\beq}{\begin{equation}}
\newcommand{\eeq}{\end{equation}}

\begin{document}
\title{The complexity of energy eigenstates as a mechanism for equilibration}

\author{Llu\'{\i}s Masanes}
\affiliation{ICFO-Institut de Ci\`encies Fot\`oniques, Mediterranean Technology Park, 08860 Castelldefels (Barcelona), Spain}

\author{Augusto J. Roncaglia}
\affiliation{ICFO-Institut de Ci\`encies Fot\`oniques, Mediterranean Technology Park, 08860 Castelldefels (Barcelona), Spain}
\affiliation{Departamento de F\'isica, FCEyN, UBA and IFIBA, CONICET,
Pabell\'on 1, Ciudad Universitaria, 1428 Buenos Aires, Argentina}

\author{Antonio Ac\'\i n}
\affiliation{ICFO-Institut de Ci\`encies Fot\`oniques, Mediterranean Technology Park, 08860 Castelldefels (Barcelona), Spain}
\affiliation{ICREA-Instituci\'o Catalana de Recerca i Estudis
Avan\c cats, Lluis Companys 23, 08010 Barcelona, Spain}

\begin{abstract}
Understanding the mechanisms responsible for the equilibration of isolated quantum many-body systems is a long-standing open problem. In this work we obtain a statistical relationship between the equilibration properties of Hamiltonians and the complexity of
their eigenvectors, provided that a conjecture about the incompressibility of quantum circuits holds. We quantify the complexity by the size
of the smallest quantum circuit mapping the local basis onto the energy eigenbasis. 
Specifically, we consider the set of all Hamiltonians having complexity $C$, and show that almost all such Hamiltonians equilibrate if $C$ is super-quadratic with the system size, which includes the fully random Hamiltonian case in the limit $C \to \infty$, and do not equilibrate if $C$ is sub-linear. We also provide a simple formula for the equilibration
time-scale in terms of the Fourier transform of the level density.
Our results are statistical and, therefore, do not apply to specific Hamiltonians. Yet, they establish a fundamental link between equilibration and complexity theory.
\end{abstract}
\maketitle

\section{Introduction}

A physical system that has been sitting for a while is often
described by a thermal or Gibbs state. This presupposes that,
whatever the initial state was, the system evolved into a
stationary state.
But this is impossible for closed systems evolving unitarily,
unless the initial state was already stationary. However, often, the reduced density matrix of a subsystem does evolve to a quasi-stationary state, and stays close to it for most of the time---this is called local equilibration. Identifying which
conditions are responsible for this process is a long-standing
question in Physics, both for classical~\cite{classic
thermo,mixing} and quantum
systems~\cite{vonNeumann,schrodinger,Gemmer}. Recently,
significant advances in the understanding of this problem have
been achieved due to three factors. First, powerful numerical
techniques have enabled the dynamical simulation of large
many-body systems~\cite{rigol,banul}. Second, the use of
quantum-information ideas, and in particular of entanglement
theory, has provided new perspectives into this
question~\cite{bristol,Cramer,gogolin}. Finally, experiments with ultra-cold atoms have allowed the manipulation and observation of many-body systems with high control~\cite{reviewultra}.
Interestingly, these experiments have challenged the current
understanding of these questions, since no thermalisation was
observed in certain non-integrable systems~\cite{weiss,schmiedmayer}.

Local equilibration can be explained by the mechanism of
dephasing~\cite{vonNeumann, Goldstein, bristol, Reimann}, under
the condition of no degenerate energy gaps (also known as no
resonant transitions)~\cite{noteengap}. In more recent
work~\cite{nou} it has been shown that if the amount of degenerate
energy gaps is small the dephasing mechanism still accounts for
local equilibration. This could be the complete answer, but it
turns out that it does not explain all equilibration processes.
For example, the systems of quasi-free bosons have infinitely-many
degenerate energy gaps, and in some cases enjoy local
equilibration~\cite{Dawson, Usha, Cramer}. Our results also allow
for constructing examples with arbitrarily many degenerate energy
gaps which still enjoy local equilibration.

In this work, we consider a different mechanism for local
equilibration, which is complementary to dephasing since it is
based on the structure of the energy eigenvectors instead of the
energy eigenvalues. This mechanism is independent of the energy
spectrum, hence it allows for constructing Hamiltonians which
enjoy local equilibration despite having many degenerate energy
gaps. Additionally, this mechanism provides a simple formula for
calculating equilibration time scales, in terms of the Fourier
transform of the level density. The time scales obtained in this
way decrease with the system's size. This is in striking contrast
with the dephasing mechanism, which provides upper bounds to the
equilibration time that grow doubly-exponentially with the
system's size~\cite{bristol, duan}. (With certain assumption on the spectrum, this  bound can be improved to just exponential in the system's size~\cite{nou}.) In the middle of these two extreme
behaviours there is the physics of extensive systems, like
Hamiltonians with local interactions, for which equilibration time
scales are expected to grow polynomially with the system's
size. This observation suggests that the mechanism introduced in
this work, based on the complexity of the energy eigenvectors, is
not responsible for the equilibration of standard extensive
systems.
However, it could account for other types of equilibration
phenomena where the underlying dynamics is sufficiently complex. Leaving aside these considerations, the main motivation of this work is purely theoretical. We consider all hermitian matrices as possible many-body Hamiltonians, without imposing any locality condition, and explore the relationship between their equilibration properties and the complexity of their eigenvectors. We believe that looking at the physics of equilibration from the perspective of quantum computation will bring new insights.

In the theory of quantum computation every algorithm can be represented by a quantum circuit, a sequence of one and two-qubit unitaries (also called gates), which processes the input data to generate the output~\cite{Nielsen}. These gates can be seen as elementary computational steps, so that harder computations require more gates. Then, it seems natural to quantify the complexity $C$ of the energy eigenvectors by the length of the smallest quantum circuit that maps the local basis onto the energy eigenbasis. In this work we establish a link between the equilibration properties of a Hamiltonian and the complexity of its eigenvectors. In order to achieve this we need a conjecture about the incompressibility of quantum circuits, also discussed in \cite{BrandaoDesigns}. Unfortunately, the mathematical techniques that are available only allow to prove a statistical relationship between equilibration and complexity.
That is, we show that the overwhelming majority of Hamiltonians with a certain complexity $C$ equilibrate or not depending on the value of $C$. However, for a given Hamiltonian with complexity $C$, we cannot be certain about its equilibration properties. Hence, our results do not provide a sharp equilibration condition, like the no degeneracy of energy gaps.
Yet, we believe that these results are important because they articulate a relationship between concepts from quantum computation and the physics of equilibration. 

The present article is structured as follows: first, in
Section~\ref{s2} we describe the physical set up that we are going
to consider throughout this work. In Section~\ref{s3} we study the
equilibration properties of Hamiltonians in the limit of large
complexity. In Section~\ref{s3} we analyse the dynamics of the
convergence towards equilibrium, and the associated time scales.
In Section~\ref{s5} we establish the relationship between the
equilibration properties of a Hamiltonian and the complexity of
its eigenstates, and formalise the above mentioned conjecture. The proofs of all our results are detailed in the
Appendix.




\section{Physical set up}
\label{s2}

We consider a system of $N$ qubits (spin-$\frac 1 2$ particles) with Hamiltonian
\begin{equation}\label{HUDU}
    H=  \sum_{n=1}^d E_n \, \proj{\Psi_n} =
    U \left[ \begin{array}{ccc}
        E_1 & & \\
        & \ddots & \\
        & & E_{d}
    \end{array} \right] U^\dagger,
\end{equation}
where $|\Psi_n \rangle$ are the eigenvectors, $E_n$ are the
eigenvalues, and $d=2^N$. The Hamiltonian characterises the
dynamics of the system: if $\rho$ is the state of the $N$ qubits
at time $t=0$ then $\rho(t) = \e^{-\I t H}\rho\, \e^{\I t H}$ is
the state at time $t$. The diagonalising unitary $U$ maps the
local basis $\ket{n} = \ket{n_1, \ldots , n_N}$ to the energy
eigenbasis $\ket{\Psi_n} = U \ket{n}$, for all $n \in \{0,1\}^N$.

Suppose that we are interested in a subset of the $N$ qubits---we
refer to it as the {\em subsystem}, while the rest of qubits are
referred to as the {\em environment}. The Hilbert space factorizes
as ${\cal H} = {\cal H}_S \otimes {\cal H}_E$, with the
corresponding dimensions satisfying $d= d_S d_E$. If $\rho$ is a
state of the $N$ qubits, the reduced state of the subsystem is
$\rho_S ={\rm tr}_E \rho$. {\em Local equilibration} happens when
the subsystem evolves towards a particular state, and stays close
to it for most of the time. If the state $\rho_S (t)$ equilibrates
to the state $\bar\rho_S$, then this must be the time-averaged
state \cite{gogolin}
\begin{equation}
    \bar\rho_S =
    \lim_{T\to \infty}\frac 1 T
    \int_0^T \hspace{-2mm} dt\, \rho_S (t)\ .
\end{equation}
As in \cite{bristol, gogolin, Cramer}, we quantify the departure
from equilibrium at time $t$ by the trace distance $\| \rho_S(t)
-\bar\rho_S \|_1$, which is directly related to the probability of
distinguishing $\rho_S(t)$ from $\bar\rho_S$ with the optimal
measurement~\cite{trace distance, Nielsen}.

In order to prove equilibration, the only condition on the
spectrum of the Hamiltonian that we need is that $g/d$ is a small
number, where $g$ is the maximal degeneracy of the Hamiltonian
\begin{equation}
    g= \max_n |\{n\rq{} : E_{n\rq{}} =E_n\}|\ .
\end{equation}
Note that this is not a very demanding condition for a large system. Apart from this, the spectrum is arbitrary, and in particular, it can have many degenerate energy gaps~\cite{noteengap}. So essentially, in our construction, the spectrum does not influence whether the Hamiltonian equilibrates or not. However, it does influence the dynamics of the convergence towards equilibrium. As we will see, this is characterised by the Fourier transform of the level density $\mu(E)$,
\begin{equation}\label{Z}
    \tilde \mu (t) \ =\ \int \!\! dE\, \mu(E)\, \e^{\I t E} \ =\ \sum_{n=1}^d \frac 1 d \  \e^{\I t
    E_n}\,,
\end{equation}
where the second equality holds for finite-dimensional systems.

\section{Equilibration in hamiltonians with generic eigenstates}\label{s3}

Before quantifying the complexity of Hamiltonians it is convenient
to study the limit of large complexity. As we see below, this
corresponds to sample the diagonalising unitary $U$ according to
the Haar measure~\cite{Haar}. The following result (proven in the
Appendix) bounds the departure from equilibrium in Hamiltonians
that are generic according to this type of sampling.

\medskip\noindent
{\bf Result 1.} For any $N$-qubit initial state $\rho$, almost all
Hamiltonians~(\ref{HUDU}) with a given spectrum $\{E_1, \ldots, E_d\}$ satisfy
\begin{equation}\label{R1}
    \| \rho_S(t) -\bar\rho_S \|_1
\leq\
    \frac{d_S^{1/2}}{\epsilon} \left(
    | \tilde \mu (t) |^4
    + \frac{g^2}{d^2}
    + \frac{7}{d_E} \right)^{\!\!1/2}
\end{equation}
for all $t$.

\medskip\noindent
The meaning of \lq\lq{}almost all\rq\rq{} is controlled by the
free parameter $\epsilon \in (0,1)$, which is an upper bound for
the fraction of Hamiltonians~(\ref{HUDU}) that violate the bound.
For example, if we set $\epsilon = 0.01$, then 99\% of the
Hamiltonians satisfy the above bound. Note that this bound is
independent of the initial state $\rho$, but the set of
Hamiltonians which violate it could depend on $\rho$. A this point
we do not know much about the nature of the Hamiltonians which
satisfy or violate this bound. In Section~\ref{s3} we are able to
say a bit more by analysing the equilibration time scales of the
Hamiltonians which satisfy the bound.

Let us discuss the significance of the three terms inside the
brackets of \eqref{R1}. The first term depends on the spectrum of
$H$ and the time $t$. At time $t=0$ the bound is useless, since
$\tilde\mu(0)=1$. For sufficiently long times, it is expected that
the phases in the sum \eqref{Z} cancel each other, resulting in a
small number, and implying equilibration. Actually, Result 2
below shows that this is the case for most of the times,
independently of the existence of degenerate energy gaps. However,
in finite systems ($d< \infty$), there are some very special (and
very long) times $t_{\rm req}$ for which $\tilde \mu (t_{\rm req})
\approx \tilde\mu(0)$. These are the quasi-recurrences, in which
the system goes back to a non-equilibrium state. In the
thermodynamic limit quasi-recurrences tend to disappear. Below, the quantity $\tilde \mu (t)$ is
calculated for some meaningful spectra. The second term also
depends on the spectrum of $H$, and implies that
Hamiltonians with huge degeneracy cannot be warranted to
equilibrate. The third term implies that equilibration
needs the environment to be much larger than the subsystem. This
condition is necessary in all approaches to equilibration known to
the authors \cite{gogolin, bristol, Reimann, Dawson, Cramer, Usha,
vonNeumann, Goldstein}.

Although quasi-recurrences take the subsystem out of equilibrium,
the following result (proven in the Appendix)
shows that, in most circumstances, the subsystem is close to the stationary state $\bar\rho_S$ for most of the time.

\medskip\noindent
{\bf Result 2.} For any $N$-qubit initial state $\rho$, almost all
Hamiltonians~(\ref{HUDU}) with a given spectrum $\{E_1, \ldots, E_d\}$ satisfy
\begin{equation}\label{R2}
    \lim_{T\to \infty}\frac 1 T \int_0^T \hspace{-3mm} dt\,
    \| \rho_S(t) - \bar\rho_S \|_1
\ \leq \
    \frac{1}{\epsilon} \left(
    \frac{g}{d_E}
    + \frac{7\, d_S}{d_E}  \right)^{\!\!1/2} .
\end{equation}

\medskip\noindent
This shows that, in the reasonable regime where $g, d_S \ll d_E$
equilibration is expected for all Hamiltonians except for a
fraction $\epsilon \in (0,1)$, when the diagonalising unitary is
sampled according to the Haar measure. This establishes the
existence of a mechanism for equilibration which is only based on
the genericness of the diagonalising unitary $U$, or
equivalently, the genericness of the energy eigenstates. This
mechanism does not rely on any condition for the energy spectrum
(other than $g\ll d_E$)---even in the presence of many degenerate
energy gaps equilibration happens.


\section{Time scales}\label{s4}

Result 1 shows that the dynamics of the convergence to
equilibrium is given by the function $|\tilde \mu (t)|$, which
only depends on the spectrum. The structure of the Fourier transform~(\ref{Z}) provides a rough time-scale for equilibration
\begin{equation}\label{seven}
    t_{\rm eq} \sim \Delta t \leq 1/\Delta E\ ,
\end{equation}
where $\Delta E$ is the variance of the distribution $\mu (E)$. In many systems of physical interest both, the range of energies $\Delta E$ and the equilibration time scale $t_{\rm eq}$, increase with the system's size. But this is incompatible with (\ref{seven}). This implies that these systems do not have a generic diagonalising unitary $U$ (according to the Haar measure), and hence, if they equilibrate, they do it by means of a different mechanism. Contrary, Hamiltonians with generic diagonalising unitaries have non-vanishing interacting terms involving any subset of the qubits constituting the system. This explains why the equilibration time scale decreases with the system size. These type of interactions, involving a large number of particles, do not happen in Nature, but can be used to articulate a connection between equilibration and complexity.


Next, we exactly calculate $|\tilde \mu (t)|$ for two standard spectra, each representing an extreme case.

\subsection{\em Spectrum of a random Hamiltonian}

Let us consider random
Hamiltonians sampled from the gaussian unitary ensemble
\cite{RandomMatrices}. According to this, each matrix element
$H_{ij} =  H_{ji}^* \in \mathbb{C}$ is an independent random
variable with probability density
\[
    P(H_{ij}) = \left\{
    \begin{array}{ll}
        (1/\pi)^{1/2}\, \e^{-H_{ij}^2} & \mbox{ if } i=j \\
        (2/\pi)^{1/2}\, \e^{-|H_{ij}|^2} & \mbox{ if } i\neq j
    \end{array} \right. .
\]
It is shown in~\cite{RandomMatrices} that, according to this
measure, the eigenvalues of $H$ are statistically independent from
the eigenvectors of $H$, and the diagonalising unitary $U$ follows
the Haar measure \cite{Haar}, as in results 1 and 2. The
gaussian unitary ensemble is often used to model some aspects of atomic nuclei,
see~\cite{nuclea rm}.

The convergence function corresponding to the spectrum of a random
matrix is, in the large-$d$ limit,
\begin{equation}
    |\tilde \mu (t)| \approx \frac{2\, J_1(t\sqrt{2d})}{\pi\, t\sqrt{2d}}\ ,
\end{equation}
where $J_1$ is the Bessel function of first kind (see Appendix).
This gives an equilibration time-scale
\begin{equation}\label{t eq rnd}
    t_{\rm eq} = \frac{1}{\sqrt{2^N}}
    \propto \frac{1}{E_{\rm max}}\ ,
\end{equation}
where $E_{\rm max}$ is the largest eigenvalue of the Hamiltonian.
Note that both, in terms of $N$ and in terms of $E_{\rm max}$, the
equilibration time \eqref{t eq rnd} is much smaller than the one
obtained for the spectrum of an integrable system, see
Eq.~\eqref{Ising t_eq} below. This is expected, since the
spectrum associated to \eqref{t eq rnd} is maximally chaotic.

\subsection{\em Spectrum of an integrable system}

Let us consider Hamiltonians with the diagonalising unitary $U$
being generic (according to the Haar measure), and the spectrum being the one of the Ising model in a transverse magnetic field $h$. Note that this is a purely academic problem, since we do not expect the Ising model to have a generic diagonalising unitary, nor an equilibration time scale that decreases with the system's size. The eigen-energies
are parametrized by the vectors $n=(n_1, \ldots ,n_N)$, where $n_k
\in \{0,1\}$ are the occupation numbers of the energy eigen-modes:
\begin{eqnarray}\label{E_n}
    E_{n} &=& \sum_{k=1}^N n_k\, \omega (2\pi k/N) ,
\\
    \omega(\phi) &=& \sqrt{(h- \cos\phi)^2 + \sin^2\phi}\ .
\end{eqnarray}
In the $t\ll 1$ regime we obtain (see Appendix)
\begin{eqnarray}\label{Z t}
    |\tilde \mu (t)|
\approx
    \e^{-t^2 N (1+h^2) /8} .
\end{eqnarray}
This gives an equilibration time-scale
\begin{equation}\label{Ising t_eq}
    t_{\rm eq} = \frac{1}{\sqrt{N (1+h^2)}}
    \propto \frac{1}{\sqrt{E_{\rm max}}} \ ,
\end{equation}
where $E_{\rm max}$ is the largest eigenvalue of the Hamiltonian
(see Appendix).




\section{Equilibration and complexity}\label{s5}

In this section we analyse the quantitative relation between equilibration and complexity, and show that as mentioned in Section~\ref{s2}, the limit of large complexity corresponds to sample the diagonalising unitary $U$ according to the Haar measure.

In the theory of quantum computation every algorithm can be represented by a quantum circuit, a sequence of one and two-qubit unitaries (also called gates), which
processes the input data to generate the output~\cite{Nielsen}.
This is analogous to classical computation, where algorithms can
be represented by circuits of logical gates. These gates can be
seen as elementary computational steps, so that harder
computations require more gates. Following this idea, 
it seems natural to quantify the complexity of the energy eigenvectors by the number of gates, denoted $C$, of the circuit that brings the local basis to the energy eigenbasis.

Now, we can obtain statistical properties of Hamiltonians as in
Results 1 and 2 above. But instead of sampling over all unitaries
(Haar measure), we sample over all unitaries that can be
implemented by a quantum circuit of a particular length $C$. It
was proven in~\cite{BrandaoDesigns} that, in this setup, taking
the limit $C \to \infty$ is equivalent to sample according to the
Haar measure. Hence, in this limit, we recover our previous
results.

\subsection{Large complexity}

A central part in the proof of Result 1 consists of performing an
average over all possible diagonalising unitaries $U$, distributed
according to the Haar measure. However, the averaged expression
only contains a fourth power of $U\otimes U^*$ (the tensor product
between $U$ and its complex-conjugate $U^*$). Here one can use the
concept of $t$-design: a finite set of unitaries $U_i \in \mathrm{SU}(d)$ with
associated probabilities $p_i$ is a $t$-design if 
\begin{equation*}
	\sum_i p_i (U_i \otimes U_i ^*)^{\otimes t} =
	\int_{\mathrm{SU}(d)} \hspace{-5mm} dU\ (U
\otimes U^*)^{\otimes t}
\end{equation*}
Then, if instead of all unitaries one averages over a subset
being a 4-design, the same result is obtained. In~\cite{viola}
strong evidence was provided to the fact that random circuits
constitute good approximations to 4-designs. Recently, this has
been rigorously proven in~\cite{BrandaoDesigns}, which allows us
to show the following.

\medskip\noindent
{\bf Result 3.} Suppose that a given initial state $\rho$ evolves
under a Hamiltonian
\begin{equation}
    H= U
    \left[ \begin{array}{ccc}
        E_1 & & \\
        & \ddots & \\
        & & E_d
    \end{array} \right]
U^\dagger\ ,
\end{equation}
where $U$ is any circuit with $C$ gates. For almost
all such circuits we have
\begin{eqnarray}\label{r3}
\nonumber &&
    \| \rho_S(t) -\bar\rho_S \|_1
\\ \label{R3} && \hspace{1cm} \leq\
    \frac{d_S^{1/2}}{\epsilon} \left(
    | \tilde \mu (t) |^4
    + \frac{g^2}{d^2}
    + \frac{7}{d_E}
    + d^3\, 2^{-\frac{\alpha C}{N}} \right)^{\!\!1/2} \qquad
\end{eqnarray}
for all $t$.

\medskip\noindent
The meaning of \lq\lq{}almost all\rq\rq{} is again controlled by the
free parameter $\epsilon \in (0,1)$, which is an upper bound for
the fraction of circuits with $C$ gates that violate the
bound~\cite{comment complexity}. The constant $\alpha$ depends on
the universal gate set, and it is calculated in \cite{viola, BrandaoDesigns}.
Compared to Result~1 there is an extra term inside the brackets in~(\ref{r3}),
which disappears in the large-$C$ limit. When $C$ is quadratic
in the number of qubits (or larger),
\begin{equation}\label{quadr}
    C \geq \alpha\rq{} N^2
\end{equation}
for $\alpha\rq{} > 3/\alpha$, the extra term is exponentially
small in $N$. One can also proceed as in the proof or Result 2, and obtain a bound for the time average of $\| \rho_S(t) -\bar\rho_S \|_1$ in the limit
$T\to\infty$. 
Note that the quadratic scaling of $C$ in~(\ref{quadr}) is the minimum needed to warrant that the circuit $U$ which diagonalises $H$ contains a gate connecting a sufficient fraction of all pairs of qubits.

In order to interpret the circuit length $C$ of $U$ as the complexity of the eigenvectors of $H$ we have to consider the following caveat. Suppose that the unitary $U$ can be written as a circuit of length $C$, and there is another unitary $U'$ which can be written as a circuit of length $C'\ll C$ and constitutes a good approximation to $U$ (e.i. the operator norm of the difference $\|U-U'\|_\infty$ is very small). In this case it does not make much sense to say that $U$ has complexity $C$. However, we conjecture that the overwhelming majority of unitaries do not have this property. Formally, for any $\epsilon>0$ and any integer $k$, the fraction of circuits of length $N^k$ which can be $\epsilon$-approximated by a circuit of length $N^{k-\alpha\epsilon}$ tends to zero as $N\to \infty$, for some positive constant $\alpha$.
Support for this conjecture is given in~\cite{BrandaoDesigns}, where the relation to equilibration is also discussed. Result~3 together with this conjecture establishes the statistical relationship between equilibration and the complexity of the energy eigenvectors.

\subsection{Small complexity}

Consider a Hamiltonian with no interaction between subsystem and
environment: \mbox{$H= H_S\! \otimes\! I_E +I_S\! \otimes\! H_E$}
where $I_S$($I_E$) is the identity matrix for the
subsystem(environment). In this case, the reduced density matrix
$\rho_S (t) = \e^{-\I t H_S} \rho_S\, \e^{\I t H_S}$ does not
converge to anything, unless it is in a stationary state from the
beginning $[\rho_S, H_S]=0$. Therefore, interaction is necessary
for equilibration. The condition of no degenerate energy gaps
implies that there is interaction across all possible bipartitions
subsystem-environment \cite{bristol}. It also implies local
equilibration, independently of the complexity of the Hamiltonian.
Therefore, in order to investigate the lack of equilibration, we
have to restrict to Hamiltonians with many degenerate energy gaps.

For simplicity, we consider Hamiltonians of the free-fermion type.
Let $\hat n_k =|1 \rangle\!\langle 1|$ be the occupation operator
for the the $k^{\rm th}$ qubit, $\omega_k$ the corresponding
excitation energy, and \beq\label{qff}
    H = U\left( \sum_{k=1}^N \omega_k \hat n_k\right) U^\dagger .
\eeq
Let the subsystem be an $M$-qubit subset of  the $N$ qubits, and the environment the remaining $N-M$ qubits. In the case $C=0$ the Hamiltonian $H= \sum_{k=1}^N \omega_k \hat n_k$ is local and its eigenvectors $|n_1, \ldots ,n_N \rangle$ are product. Hence, each qubit evolves independently, the subsystem does not interact with the environment, and there is no equilibration. Next we see that
this is still the case when the complexity $C$ is sufficiently small.

Let us lower-bound the probability that a random circuit $U$ for $N$
qubits has no gates involving any of $M$ fixed qubits. The random
circuit is generated by repeating the following process $C$ times:
uniformly pick a gate from the universal gate set; if this is a
single-qubit gate apply it to a qubit chosen uniformly from the
$N$ qubits; if this is a two-qubit gate  apply it to a pair of
qubits chosen uniformly. The probability $p$ that no gate is applied
to any of the $M$ qubits satisfies
\beq
    p \ \geq\  \left(\frac{N-M}{N} \right)^{2C} .
\eeq
In this event, there is no interaction subsystem-environment, and hence, no
equilibration. Suppose the complexity is sublinear: $C \leq
N^{\nu}$ with $0< \nu <1$. If we fix the size of the subsystem
$M$, in the large-$N$ limit we have $p \approx 1- 2M/N^{1-\nu}$,
and then

\medskip\noindent {\bf Result 4.} For almost all circuits $U$ with
sublinear complexity, the associated Hamiltonian
\eqref{qff} does not enjoy local equilibration.

\section{Conclusions}

In this work we have addressed the problem of equilibration in
isolated quantum many-body systems evolving under unitary
dynamics. We have pointed out the existence of a mechanism for
local equilibration which is based on the complexity of the energy
eigenvectors. We have shown that: almost all Hamiltonians whose diagonalising unitary is a circuit of length $C$ equilibrate if $C$ is super-quadratic with the
system's size, and do not equilibrate if $C$ is sub-linear with
the system's size. What happens in between these two regimes is an
open problem that we leave for the future. Since these results are
statistical, it is difficult to extract conclusions for physically
relevant Hamiltonians, like those with local interactions.

Under the action of this equilibration mechanism, the
equilibration time scale decreases with the system's size. This is
not expected in Hamiltonians with local interactions. In particular, the equilibration time scale of the Ising model with long-range interactions \cite{Kastner} diverges with the system's size.
Clearly, this class of Hamiltonians belong to the
$\epsilon$-fraction of cases that violate our bounds.
However, our results could apply to sufficiently chaotic systems.

The relation between equilibration and the complexity of solving the dynamics of a physical system that emerges from our
results resembles the situation in classical mechanics, where the notion of integrability plays an important role~\cite{mixing}. In
fact, there exists a link between equilibration (formalised by
weak mixing~\cite{mixing}) and the difficulty of solving the
dynamics of a classical system: integrable systems violate weak mixing, while sufficiently chaotic systems satisfy it. Now, if the circuit size is interpreted as the complexity of solving the
dynamics of a quantum system, the resulting picture resembles what
happens in classical mechanics. Hence, our results may
also contribute to the problem of finding a definition
for quantum integrability---a proposal in terms of computational
complexity can be found in~\cite{qintegr}.





Dephasing under the condition of no degenerate energy gaps
\cite{bristol,Reimann} and the complexity of the energy
eigenvectors are two independent mechanisms that explain the
phenomenon of local equilibration. Are there other mechanisms,
apart from these two? Does any of them play a dominant role in
natural phenomena?

\medskip

{\em Note added after completion of this work:} results related to
the ones presented here have been obtained independently in
References~\cite{Vinayak} and~\cite{Brandao}.

\acknowledgments

We are very thankful to Fernando G. S.
L. Brand\~ao, Ignacio Cirac, Michal Horodecki, Maciej Lewenstein,
Anthony J. Short for discussions on the topic of this work. This
work is financially supported by the ERC Starting Grant PERCENT,
the EU FP7 Q-Essence and Spanish FIS2010-14830 projects,
CatalunyaCaixa and the Generalitat de Catalunya. AR acknowledges support
from CONICET.

\newpage
\begin{appendix}

\section{Equilibration bounds}\label{app:bounds}

\subsection{Proof of Result 1}

Consider the linear map
\begin{equation}\label{identity 1}
    \Omega_t [\rho]=
    \sum_{E_n\neq E_{n\rq{}}} \hspace{-2mm}
    \e^{\I t(E_{n} -E_{n\rq{}})}
    {\rm tr}_E (\Psi_n \rho \Psi_{n\rq{}})\  ,
\end{equation}
and note that $\Omega_t [\rho] = \rho_S(t) -\bar\rho_S$. The sum $\sum_{E_n\neq E_{n\rq{}}}$ runs over all pairs of eigenstates $n, n\rq{} \in \{1,\ldots , d\}$ with different energies $E_n\neq E_{n\rq{}}$. Any $d \times d$ matrix $B$ satisfies $\|B\|_1 \leq \sqrt{d}\, \|B\|_2$ where $\| B \|_2 = \sqrt{{\rm tr}(B^\dagger B)}$; hence we have the bound
\begin{equation}\label{norm 1 2}
    \| \rho_S (t) -\bar\rho_S \|_1 \ \leq\
    \sqrt{d_S    {\rm tr}_S (\Omega_t [\rho]^2)}\ .
\end{equation}
Let $\{ \ket{s}; s=1,\ldots, d_S\}$ be an orthonormal basis of ${\cal H}_S$, and $\{ \ket{e}; e=1,\ldots, d_E\}$ an orthonormal basis of ${\cal H}_E$.

Before doing the general case, we first consider the case where the initial state is pure $\psi = \proj{\psi}$. Some calculation shows
\begin{eqnarray}\label{chain1}
&&
    {\rm tr}_S (\Omega_t [\psi]^2)
    \phantom{\sum}
\\ \nonumber &=&
    \sum_{E_n\neq E_{n^\prime}} \sum_{E_k \neq E_{k\rq{}}}
    \e^{\I t(E_n -E_{n\rq{}} + E_k - E_{k\rq{}})}
\\ \nonumber &\times&
    \sum_{s,e,s\rq{}, e\rq{}}
    \bra{ se} \Psi_n \psi \Psi_{n\rq{}} \ket{s\rq{} e}\!
    \bra{s\rq{} e\rq{}} \Psi_k \psi \Psi_{k\rq{}} \ket{s e\rq{}}
\\ \nonumber &=&
    \sum_{E_n\neq E_{n^\prime}} \sum_{E_k \neq E_{k\rq{}}}
    \e^{\I t(E_n -E_{n\rq{}} + E_k - E_{k\rq{}})}
\\ \nonumber &\times&
    \bra{ n,k,n^\prime,k^\prime} (U^\dagger)^{\otimes 4} M U^{\otimes 4} \ket{n^\prime,k^\prime, n,k}
\end{eqnarray}
where
\begin{equation}\label{M}
M=  \sum_{s,e,s',e'} \psi \otimes \psi \otimes
\ket{e s}\!\bra{e s'}\otimes \ket{e' s'}\!\bra{e' s}\ .
\end{equation}
We want to calculate the average of ${\rm tr}_S (\Omega_t [\psi]^2)$ over all unitaries from ${\rm SU}(d)$ according to the Haar measure~\cite{Haar}. To do this we first compute
\begin{equation}\label{lrU}
    M_0= \Big\langle (U^\dagger)^{\otimes 4} M U^{\otimes 4}  \Big\rangle_U =
    \int_{{\rm SU}(d)}\hspace{-5mm} dU\ (U^\dagger)^{\otimes 4} M U^{\otimes 4} \ .
\end{equation}
Due to the Schur-Weyl duality~\cite{FH}, the matrix $M_0$ is a linear combination of permutations
\begin{equation}\label{e6}
M_0=\sum_\pi c_\pi V_\pi\ ,
\end{equation}
where the index $\pi$ runs over the 4! permutations of four elements, $c_\pi \in \mathbb{C}$ are some coefficients, and the unitaries $V_\pi$ permute the four factor spaces in which $M$ acts. For instance
$\bra{n_1, n_2, n_3, n_4} V_{(2341)}= \bra{n_2, n_3, n_4, n_1}$.

Let us obtain the coefficients $c_\pi$ from~(\ref{e6}). Note that $M_0$ has the following symmetries
\[
M_H =M_H V_{(2134)}=V_{(2134)}M_H=V_{(1243)}M_HV_{(1243)}\ ,
\]
which implies the following identities
\begin{eqnarray}
    c_1 &:=& c_{(1234)} =  c_{(2134)} \ , \\
    c_2 &:=& c_{(1243)} = c_{(2143)} \ , \nonumber\\
    c_3 &:=& c_{(1423)}= c_{(1342)}= c_{(2413)}= c_{(4123)} \nonumber\\
    &=&  c_{(4213)}= c_{(2341)}= c_{(3142)}= c_{(3241)} \ , \nonumber\\
    c_4 &:=& c_{(1324)}= c_{(1432)}= c_{(2314)}= c_{(3124)} \nonumber\\
    &=&  c_{(3214)}= c_{(2431)}= c_{(4132)}= c_{(4231)} \nonumber\\
    c_5 &:=& c_{(3412)}= c_{(4321)}= c_{(3421)}= c_{(4312)} \ .\nonumber
\end{eqnarray}
These four different coefficients can be determined with the following equations:
\begin{eqnarray}\label{e8}
&& {\rm tr}(M_0 V_{(1234)})=d\, d_E\ , \\
&& {\rm tr}(M_0 V_{(1243)})=d^2/d_E\ , \nonumber\\
&& {\rm tr}(M_0 V_{(1342)})=d /d_E\ , \nonumber\\
&& {\rm tr}(M_0 V_{(1324)})=d_E\ , \nonumber\\
&& {\rm tr}(M_0 V_{(4312)})= {\rm tr}_S ({\rm tr}_E^2 \proj{\psi}) =: \beta \ . \nonumber
\end{eqnarray}
These equations follow from the identity ${\rm tr} V_\pi = d^{\rm{cicl}(\pi)}$, where $\rm{cicl}(\pi)$ is the number of cicles in the permutation $\pi$. The solution of the system of equations~(\ref{e8}) is:
\begin{eqnarray*}
&& c_1=\frac{d (d (d+4)+2) \left(d_E^2-1\right)-2 d_E^2+2 d_E  \beta}{(d-1) d^2 (d+1) (d+2) (d+3) d_E}\ , \\
&& c_2=\frac{d \left(d^3+4 d^2-(d+4) d_E^2+2 d-2\right)-2 d_E^2+2 d_E \beta}{(d-1) d^2 (d+1) (d+2) (d+3) d_E}\ , \\
&& c_3=\frac{-(d+1) d_E \beta +d+d_E^2}{(d-1) d^2 (d+1) (d+2) (d+3) d_E}\ , \\
&& c_4=\frac{-(d+1) d_E \beta+d+d_E ^2}{(d-1) d^2 (d+1) (d+2) (d+3) d_E }\ , \\
&& c_5=\frac{d (d_E \beta -1)+d_E ( \beta -d_E)}{d^2 \left(d^3+3 d^2-d-3\right) d_E}\ .
\end{eqnarray*}
Assuming $d>d_E >0$ and using $\beta \leq 1$ we obtain
\begin{eqnarray*}
    |c_2| & \leq&  \frac{d \left(d^3+4 d^2+2 d\right)+2 d_E}{(d-1) d^2 (d+1) (d+2) (d+3) d_E} \\
&\leq& \frac{\left(d^2+4 d+4 \right)}{(d-1) (d+1) (d+2) (d+3) d_E} \\
&\leq& \frac{d+2}{(d-1) d^2 d_E} \leq  \frac{2}{d^2 d_E}\ , \\
    |c_3| &\leq& \frac{d+d\,d_E}{(d-1) d^4 (d_E+1)  d_E}
\leq  \frac{1}{d^3 d_E^2 }\ ,  \\
    |c_5| &\leq& \frac{d \, d_E+d}{d^5  d_E} \leq \frac{1+d_E}{d^4\, d_E }\ .
\end{eqnarray*}

Combining~(\ref{chain1}) and~(\ref{e6}) we get
\begin{equation}
    \left\langle {\rm tr}_S (\Omega_t [\psi]^2)\right\rangle_U
   = \sum_\pi c_\pi f_\pi (t)\ ,
\end{equation}
where
\begin{eqnarray}
    f_\pi (t) &=& \sum_{E_n\neq E_{n^\prime}} \sum_{E_k \neq E_{k\rq{}}}
    \e^{\I t(E_n -E_{n\rq{}} + E_k - E_{k\rq{}})} \\
     &\times& \bra{ n,k,n^\prime,k^\prime} V_\pi \ket{n^\prime,k^\prime, n,k}\ . \nonumber
\end{eqnarray}
The constraints $E_n\neq E_{n^\prime}$ and $E_k \neq E_{k\rq{}}$ imply that $f_\pi(t)=0$ for most $\pi$. The only permutations $\pi$ for which $f_\pi(t) \neq 0$ are following ones
\begin{eqnarray*}
    f_{(2143)} (t) &=&
    \sum_{E_n\neq E_{n^\prime}} 1\ ,
\\
    f_{(2413)} (t)  &=&
    \sum_{E_{n} \neq E_{n\rq{}} \neq E_{k\rq{}} } \hspace{-4mm}
    \e^{\I t(E_n - E_{k\rq{}})}\ ,
\\
    f_{(3142)} (t)  &=&
    \sum_{E_{k} \neq E_n \neq E_{n\rq{}} } \hspace{-4mm}
    \e^{\I t(E_k - E_{n\rq{}})}\ ,
\\
    f_{(3412)}(t)  &=&
        \!\! \sum_{E_n\neq E_{n^\prime}} \sum_{E_k \neq E_{k\rq{}}}
    \!\!  \e^{\I t(E_n -E_{n\rq{}} + E_k - E_{k\rq{}})}\ ,
\\
        f_{(4321)}(t) &=&
     \sum_{E_n\neq E_{n^\prime}}
    \e^{\I t2( E_n - E_{n\rq{}})}\ ,
\\
    f_{(4312)} (t)&=&
    \sum_{E_n\neq E_{n^\prime} \neq E_k} \hspace{-4mm}
     \e^{\I t(E_n +E_k - 2 E_{n\rq{}})}\ ,
\\
    f_{(3421)} (t)  &=&
    \sum_{E_{n\rq{}} \neq E_n \neq E_{k\rq{}} } \hspace{-4mm}
    \e^{\I t(2 E_n -E_{n\rq{}} - E_{k\rq{}})}\ .
\end{eqnarray*}
In summary:
\begin{eqnarray}\label{summ}
&&
    \left\langle {\rm tr}_S (\Omega_t [\psi]^2)\right\rangle_U
\\ \nonumber
    &=& c_2 f_{(2143)}(t)+c_3 \left[ f_{(2413)}(t)+f_{(3142)} (t)\right] +
\\ \nonumber
    &+& c_5 \left[ f_{(3412)}(t)+f_{(4321)}(t)+f_{(4312)}(t)+f_{(3421)} (t)\right].
\end{eqnarray}
Using
\begin{equation*}
    \sum_{E_n \neq E_k } 1 \leq d (d-1) \quad \mbox{ and }
    \sum_{E_{n\rq{}} \neq E_n \neq E_{k\rq{}} } \hspace{-4mm} 1 \leq d^2 (d-1)
\end{equation*}
we obtain
\begin{eqnarray}\label{2e}
    \left| c_2 f_{(2143)}(t) \right| &\leq& \frac{2}{d_E}\ ,
\\ \label{2d2}
    \left| c_3 \left[ f_{(2413)}(t)+f_{(3142)} (t)\right] \right| &\leq& \frac{2}{d_E^2}\ ,
\end{eqnarray}
and also
\begin{eqnarray}
&&  \left| c_5 \left[ f_{(4321)}(t)+f_{(4312)}(t)+f_{(3421)} (t) \right] \right|
\nonumber \\
    &\leq& \frac{1+d_E}{d^4\, d_E } \left[ (d^2-d) + 2d (d^2-d)\right]
\nonumber \\ \label{2de}
    &\leq& \frac{2d^3-d^2-d}{d^3\, d_E } \leq \frac{2}{d_E}\ .
\end{eqnarray}
Define $w= (\sum_{E_n = E_{n^\prime}} \! 1)$ and note that $w\leq gd$. Direct calculation shows
\begin{eqnarray}
\nonumber &&
    \left| c_5\,  f_{(3412)} (t) \right|
    \phantom{\sum}
\\ \label{interest} &\leq&
    \sum_{E_n\neq E_{n^\prime}} \sum_{E_k \neq E_{k\rq{}}}
    d^{-4}\, \e^{\I t(E_n -E_{n\rq{}} + E_k - E_{k\rq{}})}+d_E^{-1}\quad
\\ \nonumber &=&
    \left| \tilde \mu (t) \right|^4 + w^2 d^{-4}
    - 2\,w\, d^{-2}  \left| \tilde \mu (t) \right|^2 +d_E^{-1}\phantom{\sum}
\\ \label{eq22} &\leq&
    \left| \tilde \mu (t) \right|^4 + g^2 d^{-2}+d_E^{-1} .
    \phantom{\sum}
\end{eqnarray}
Substituting~(\ref{2e}-\ref{eq22}) in~(\ref{summ}) gives
\begin{equation}\label{E17}
    \left\langle {\rm tr}_S (\Omega_t [\psi]^2)\right\rangle_U
\ \leq\
    \left| \tilde \mu (t) \right|^4 + \frac{g^2}{d^2} + \frac{7}{d_E}\ .
\end{equation}

Let $\rho$ be the not-necessarily-pure initial state ($\rho = \sum_i p_i \psi_i$ where each $\psi_i$ is pure). Any real-valued random variable $X$ satisfies $\langle X \rangle \leq \langle X^2 \rangle^{1/2}$. Using this, the triangular inequality, and \eq{norm 1 2}, we obtain
\begin{eqnarray}
\nonumber &&
    \Big\langle \| \rho_S (t) -\bar\rho_S \|_1 \Big\rangle_{U}
\\ \nonumber &=&
    \Big\langle \| \sum_i p_i\, \Omega_t [\psi_i] \|_1 \Big\rangle_{U}
\\ \nonumber &\leq&
    \sum_i p_i\, \Big\langle \| \Omega_t [\psi_i] \|_1 \Big\rangle_{U}
\\ \nonumber &\leq&
    \sum_i p_i\, \Big\langle d_S\,
    {\rm tr}_S (\Omega [\psi_i]^2)
    \Big\rangle_{U}^{1/2}
\\  \label{eq23} &\leq&
    d_S^{1/2} \Big(\left| \tilde \mu (t) \right|^4 + \frac{g^2}{d^2} + \frac{7}{d_E} \Big)^{1/2} \ .
\end{eqnarray}

We conclude the proof of Result 1 with a simple probabilistic argument. Let $X$ be a
random variable taking positive values such that $\langle X \rangle \leq x_0$. If $\epsilon = {\rm prob}\{X > x\}$ then $(1-\epsilon)0 + \epsilon\, x \leq \langle X \rangle$,
therefore $x\leq x_0/\epsilon$.

\subsection{Proof of Result 2}

If expression~(\ref{interest}) instead of~(\ref{eq22}) is used in the chain of inequalities~(\ref{eq23}) then one obtains the following.
\begin{eqnarray}\label{R12}
&&
    \| \rho_S(t) -\bar\rho_S \|_1
\\ \nonumber &\leq&
    \frac{d_S^{1/2}}{\epsilon}\! \left(
    \sum_{E_n\neq E_{n^\prime}} \sum_{E_k \neq E_{k\rq{}}} \!\!
    d^{-4}\, \e^{\I t(E_n -E_{n\rq{}} + E_k - E_{k\rq{}})}
    + \frac{7}{d_E} \right)^{\!\!1/2}
\end{eqnarray}
Note that
\begin{eqnarray}
&&
    \lim_{T\to \infty} \frac 1 T \int_0^T \hspace{-2mm} dt\,
    \e^{\I t(E_n -E_{n\rq{}} + E_k - E_{k\rq{}})}
\\ \nonumber &=&
    \sum_{E_n\neq E_{n^\prime}} \sum_{E_k \neq E_{k\rq{}}} d^{-4}
    \delta(E_n -E_{n\rq{}} + E_k - E_{k\rq{}})
    \ \leq\ \frac g d \ ,
\end{eqnarray}
where $\delta(E_n -E_{n\rq{}} + E_k - E_{k\rq{}})$ is a Kronecker delta.
This bound, inequality \eqref{R12}, and the convexity of the square root, imply
\begin{equation}
    \lim_{T\to \infty} \frac 1 T \int_0^T \hspace{-2mm} dt\,
    \| \rho_S(t) - \bar\rho_S \|_1
\leq
    \frac{1}{\epsilon} \sqrt{
     \frac{g}{d_E} + \frac{7\, d_S}{d_E} } .
\end{equation}
This shows Result 2.

\subsection{Proof of Result 3}

Result 3 can be proven by following the same steps as in the proof of Result 1, but replacing the average over the Haar measure by the average over unitaries which are circuits with $C$ gates. As in (\ref{lrU}) we define the linear map
\[
    {\cal E} (X) =
    \Big\langle (U^\dagger)^{\otimes 4} X U^{\otimes 4}  \Big\rangle_U
\]
which symmetrises any $d^4 \times d^4$ matrix $X$. It can be shown that ${\cal E}$ is a projector~\cite{FH}, so its eigenvalues are 1 and 0. Let ${\cal K}(C)$ be the set of $N$-qubit circuits with $C$ gates, from a particular universal gate set. We denote by $\langle \cdot \rangle_{U\in {\cal K}(C)}$ the average over all unitaries in ${\cal K}(C)$ with equal weights. It is shown in~\cite{viola} that
\begin{equation}\label{E}
    \Big\langle (U^\dagger)^{\otimes 4} X U^{\otimes 4}  \Big\rangle_{U\in {\cal K}(C)} =
    {\cal E}(X) +\lambda^C {\cal E}\rq{} (X),
\end{equation}
where $\lambda = 1- \alpha/N$, the constant $\alpha>0$ depends on the universal gate set, and the linear map ${\cal E}\rq{}$ has bounded norm $\|{\cal E}\rq{} (X)\|_2 \leq 1$ for all $X$ with $\|X\|_2 = {\rm tr}(X^\dagger X) \leq 1$.
The matrix $M$ defined in (\ref{M}) satisfies   $\| M \|_2 = d_E \sqrt{d_S}$. Using this, identity (\ref{E}) and the fact that $\sum_{E_n\neq E_{n^\prime}} \sum_{E_k \neq E_{k\rq{}}} 1 \leq d^4$, we have
\begin{eqnarray}\label{ch2}
&&
    \left| \left\langle
        {\rm tr}_S (\Omega_t [\psi]^2)
    \right\rangle_{U\in {\cal K}(C)} \right|
    \phantom{\sum}
\\ \nonumber &\leq&
    \left| \left\langle
        {\rm tr}_S (\Omega_t [\psi]^2)
    \right\rangle_U \right|
\\ \nonumber &+&
    \sum_{E_n\neq E_{n^\prime}} \sum_{E_k \neq E_{k\rq{}}}
    \left| \bra{ n,k,n^\prime,k^\prime}
    {\cal E}\rq{}(M) \ket{n^\prime,k^\prime, n,k} \right|
\\ \nonumber &\leq&
    \left| \tilde \mu (t) \right|^4 + \frac{g^2}{d^2} + \frac{7}{d_E} +
    d^4 \lambda^C d_E \sqrt{d_S}\ .
\end{eqnarray}
Reproducing the argument of Result~1, but using (\ref{ch2}) instead of (\ref{E17}) one obtains Result~3.

\section{Calculation of $|\tilde\mu (t)|$}
\label{app:time_scales}


\subsection{Spectrum of a random matrix}

In this section we calculate the convergence function $|\tilde \mu (t)|$ for the spectrum of a random Hamiltonian $H$, generated by the probability distribution $P(H)$ corresponding to the gaussian unitary ensemble \cite{RandomMatrices}. According to this, each matrix element $H_{ij}$ is an independent
random variable; the elements in the diagonal $H_{ii} \in \mathbb{R}$ have probability density
$P(H_{ii}) = \pi^{-1/2}\, \e^{-H_{ii}^2}$; the elements not in the diagonal
$H_{ij} = \overline H_{ji} \in \mathbb{C}$ have probability density
$P(H_{ij}) = (2/\pi)\, \e^{-2|H_{ij}|^2}$. This is equivalent to say that the diagonalising unitary
$U$ of $H$ follows the uniform distribution over unitaries (the Haar measure \cite{Haar}), and independently,
the spectrum of $H$ follows the probability density
\beq
P(E_1, \ldots ,E_d) = \alpha\, \e^{-\sum_{i=1}^d E_i^2} \!\! \prod_{1 \leq i < j \leq d} \!\! |E_i -E_j|^2
\eeq
where $E_i \in (-\infty, \infty)$, and $\alpha$ is a normalisation constant. That is, eigenvalues and eigenvectors
are independent random variables.

Since, what appears in Result~1 is $|\tilde \mu (t)|$ to the fourth power, we are going to calculate the average
\beq\nonumber
    \left\langle |\tilde \mu (t)|^4 \right\rangle_H =
    \int\!\! dE_1 \cdots dE_d\, P(E_1, \ldots ,E_d)\, |\tilde \mu (t)|^4.
\eeq
A standard trick within random matrix theory is that, with probability almost one, the value of $|\tilde \mu (t)|^4$ for a randomly chosen spectrum is very close to the above average.

For the next, it is useful to define the $n$-point correlation-function
\[
    R_n (E_1, ... , E_n) =
    \frac{d!}{(d-n)!} \int\!\! dE_{n+1} \cdots dE_d\, P(E_1, \ldots ,E_d)
\]
where $0<n<d$ \cite{RandomMatrices}. The sum 
\[
    d^4 \left\langle |\tilde \mu (t)|^4 \right\rangle_H =
    \sum_{ijkl} \left\langle \e^{\I t(E_i-E_j+E_k-E_l)} \right\rangle_H
\]
can be split in the four terms where $i,j,k,l$ are: (i) all different, (ii) two of them equal;
(iii) three of them equal or two pairs equal, and (iv) all equal. These four terms are separated in
\begin{widetext}
\beqa
    d^4 \left\langle |\tilde \mu (t)|^4 \right\rangle_H
    &=&
    \int\!\! dE_1\, dE_2\, dE_3\, dE_4\ R_4 (E_1, E_2, E_3 ,E_4)\
    \e^{\I t(E_1-E_2+E_3-E_4)}
\nonumber \\ &+&
    \int\!\! dE_1\, dE_2\, dE_3\ R_3 (E_1, E_2, E_3)
    \left[ \e^{\I t(2 E_1-E_2-E_3)} +\e^{-\I t (2E_1-E_2-E_3)}
    +4\, \e^{\I t (E_1-E_2)}
    \right]
\nonumber \\ &+&
    \int\!\! dE_1\, dE_2\ R_2 (E_1, E_2)
    \left[ \e^{\I t 2(E_1-E_2)}+4\, \e^{\I t (E_1-E_2)} \right]
    \  + \ \left[ 2d^2-d \right].
\label{eq:GUEbound}
\eeqa
\end{widetext}
It is shown in \cite{RandomMatrices} that the $n$-point correlation functions can be written as
\beq
R_n (E_1, \ldots , E_n)= \det[K(E_i, E_j)]_{i,j=1,...,n}
\label{eq:det}
\eeq
where
\beq
K (E_i, E_j)= \sum_{k=0}^{d-1} \varphi_k (E_i)\, \varphi_k (E_j)\ ,
\label{eq:Ks}
\eeq
where $\varphi_k (x)$ are the eigenfunctions of the quantum harmonic oscillator. To see how the determinant
works in \eq{eq:det} consider the example
\[
    R_2 (E_1, E_2) = K(E_1, E_1)\, K(E_2, E_2) - K(E_1, E_2)^2 \ .
\]
When substituting \eqref{eq:det} in \eqref{eq:GUEbound}, we obtain products of objects of the form
\beq
    \int\!\! dE\, K(E, E)\, \e^{\I t E}
\label{eq:FTDensity}
\eeq
and
\beqa
    \int\!\! dE_1 \cdots dE_n\, K(E_1, E_2) \cdots K(E_n, E_1)\,
    \e^{\I (t_1 E_1+\cdots + t_n E_n)}
\nonumber \\ \label{eq:FTK} =
    {\rm tr}\! \left[ P\, \e^{\I t_1  X}\,  P\, \e^{\I t_2  X}\,
    \cdots  P\, \e^{\I t_n  X} \right]
    \hspace{3cm}
\eeqa
for $n=2,3,4$---where $ P=\sum_{k=0}^{d-1} \ket{\varphi_k}\! \bra{\varphi_k}$ is the projector
onto the $d$-dimensional lower-energy subspace of the harmonic oscillator, and $ X$ is the position
operator. For any pair of bounded operators $A, B$, the Cauchy-Schwartz inequality tells
\beq
    \big| {\rm tr}[A^\dagger B ] \big|
    \leq
    \sqrt{
        {\rm tr}[A^\dagger A ]\,
        {\rm tr}[B^\dagger B ]
    }\ .
\eeq
This can be used to bound \eqref{eq:FTK} for the case $n=4$:
\beqa
\nonumber
&&
    \Big| {\rm tr} \big[
        \big(P\, \e^{\I t_1 X} P\, \e^{\I t_2 X} \big)
        \big(P\, \e^{\I t_3 X} P\, \e^{\I t_4 X} \big)
    \big] \Big|
\\ \nonumber &\leq &
    \sqrt{{\rm tr} \big[\e^{-\I t_2 \hat X} \hat P\, \e^{-\I t_1 \hat X} \hat P^2 \,
    \e^{\I t_1 \hat X} \hat P\, \e^{\I t_2 \hat X} \big]
    {\rm tr} \big[\hat P\, \e^{-\I t_3 \hat X} \hat P\, \e^{\I t_3 \hat X} \big]}
\\ \label{bn4} &\leq &
    \sqrt{
        \sqrt{ {\rm tr} \big[\hat P \big] {\rm tr} \big[\hat P \big] }
        \sqrt{ {\rm tr} \big[\hat P \big] {\rm tr} \big[\hat P \big] }
    }
\leq  d.
\eeqa
By setting $t_4=0$ and $t_3 =0$, we obtain the same bound for $n=3,2$. It is shown in
\cite{RandomMatrices} that in the large-$d$ limit we have
\beq\label{semi-circle}
    K (E,E) \approx \left\{
    \begin{array}{ccc}
        \frac 1 \pi \sqrt{2d-E^2} & \mbox{ if } & |E| \leq \sqrt{2d} \\
        0 & \mbox{ if } & |E| > \sqrt{2d} \\
    \end{array} \right.\ .
\eeq
This is called \lq\lq{}the semi-circle law\rq\rq{}. In this limit, the integral \eqref{eq:FTDensity} can be evaluated
\beq\label{J}
    \int\!\! dE\, K(E, E)\, \e^{\I t E}
    \ \approx\
    d\, \frac{2\, J_1(t\sqrt{2d})}{t\sqrt{2d}} \ \leq\ d\  ,
\eeq
where $J_1$ is the Bessel function of the first kind. When substituting \eqref{eq:det} in \eqref{eq:GUEbound},
we obtain several terms, each being a product of objects of the form \eqref{eq:FTDensity} or \eqref{eq:FTK}.
According to \eqref{bn4} and \eqref{J}, each of these factors (\ref{eq:FTDensity}) or (\ref{eq:FTK}) is
bounded by $d$. Since $\left\langle |\tilde \mu (t)|^4 \right\rangle_H$ is equal to \eqref{eq:GUEbound}
divided by $d^4$, all terms are of order $1/d$ or smaller, except for the single term with a four-fold
product of \eqref{eq:FTDensity}. This implies that in the large-$d$ limit we have
\beq
    \left\langle |\tilde \mu (t)|^4 \right\rangle_H \ \approx\
    \left( \frac{2\, J_1(t\sqrt{2d})}{t\sqrt{2d}} \right)^4 .
\eeq

\subsection{Spectrum of the Ising model}

The eigen-energies are parametrized by the vectors
$n=(n_1, \ldots ,n_N)$, where $n_k \in \{0,1\}$ are the occupation numbers of the energy eigen-modes:
\begin{eqnarray}
    E_{n} &=& \sum_{k=1}^N n_k\, \omega (2\pi k/N) ,
\\
    \omega(\phi) &=& \sqrt{(h- \cos\phi)^2 + \sin^2\phi}\ .
\end{eqnarray}
Then
\begin{eqnarray}
\nonumber
    |\tilde \mu (t)|^2 &=&
    2^{-2 N} \prod_{k=1}^N \left| \sum_{n_k=0}^1
    \e^{\I t n_k \omega (2\pi k/N)}\right|^2
\\ \nonumber &=&
    2^{-N} \prod_{k=1}^N \left( 1 + \cos[t \omega (2\pi k/N)] \right)
\\ \label{eq9} &\approx&
    2^{-N} \exp\!\left[ \frac{N}{2\pi} \int_0^{2\pi} \hspace{-4mm} d\phi\,
    \ln\! \left( 1 + \cos[t\, \omega (\phi)]\right) \right] , \qquad
\end{eqnarray}
where the approximation holds for large $N$. Let us analyse this expression in the small-$t$ and large-$t$ limits.
For $t\ll 1$ we have, up to fourth-order terms,
\begin{eqnarray}\nonumber
    \ln\! \left( 1 + \cos[t\, \omega (\phi)]\right)
 \approx
    \ln 2 - \left( 1+h^2 -2h \cos[\phi]\right) \frac{t^2}{4} ,
\end{eqnarray}
which gives
\begin{eqnarray}
    |\tilde \mu (t)|^2
\approx
    \e^{-t^2 N (1+h^2) /4} .
\end{eqnarray}
For $t\gg 1$ we have
\begin{eqnarray}
\nonumber
    |\tilde \mu (t)|^2
&\leq&
    2^{-N} \exp\!\left[ N
    \ln\! \left( 1 + \frac{1}{2\pi} \int_0^{2\pi} \hspace{-4mm} d\phi\,
    \cos[t\, \omega (\phi)]\right) \right]
\\ \label{eq10} &\approx&
    2^{-N} \exp\!\left[ N
    \ln(1)\right]
=
    2^{-N},
\end{eqnarray}
where the inequality follows from the convexity of the logarithm, and the approximation holds when the integrand oscillates heavily, that is when $t\gg 1$. In the large-$N$ limit, the largest eigenvalue is
\begin{equation}
    E_{\rm max} =
    \frac{N}{2\pi} \int_0^{2\pi} \hspace{-4mm} d\phi\,
    \sqrt{(h- \cos\phi)^2 + \sin^2\phi}\ ,
\end{equation}
which is proportional to $N$.
\end{appendix}


\begin{thebibliography}{99}
\bibitem{noteengap} A Hamiltonian has no degenerate gap
whenever its spectrum is such that if $E_m-E_n=E_{m'}-E_{n'}$ then
$m=n$ and $m'=n'$, or, $m=m'$ and $n=n'$.
\bibitem{trace distance} The trace-norm of a matrix $A$ is $\|A\|_1 = {\rm tr} \sqrt{A^\dagger A}$. The probability of discriminating between $\rho_1$ and $\rho_2$ is $p= \frac 1 2 + \frac 1 4 \|\rho_1 - \rho_2\|_1$.
\bibitem{comment complexity} Recall that the number of circuits with $C$ gates is finite, hence there is no need to define a measure over them. Also, different circuits can give rise to the same unitary or Hamiltonian, hence, the bound on the proportion $\epsilon$ cannot be directly translated to the set of hamiltonians.
\end{thebibliography}
\end{document}